\documentclass[11pt,a4paper]{article}
\usepackage{amsmath,amssymb}
\usepackage{epsfig,graphicx}
\usepackage{cite}

\def\mprp{\mbox{\tiny $\bot$}}
\def\mprl{\mbox{\tiny $\|$}}

\def\D{\mathrm{d}} 

\def\beq{\begin{eqnarray}}
\def\eeq{\end{eqnarray}}
\def\ee{\varepsilon}
\def\lm{\lambda}

\newcommand{\prl}[1]{#1_{\mbox{\tiny $\|$}}}

\def\P{{\cal P}}

\begin{document}

\title{\vspace*{-20mm}
\begin{flushright}
{\normalsize Yaroslavl State University\\
             Preprint YARU-HE-08/04\\
             hep-ph/yymmnnn} \\[10mm]
\end{flushright}
\bf Influence of the photon - neutrino processes on magnetar cooling}
\author{M.~V.~Chistyakov$^a$\footnote{{\bf e-mail}: mch@uniyar.ac.ru},
D.~A.~Rumyantsev$^{a}$\footnote{{\bf e-mail}: rda@uniyar.ac.ru}
\\
$^a$ \small{\em Division of Theoretical Physics, Department of Physics, 
Yaroslavl State University} \\
\small{\em Sovietskaya 14, 150000 Yaroslavl, Russia}
}
\date{}
\maketitle

\begin{abstract}
The photon-neutrino processes $\gamma e^{\pm} \to e^{\pm} \nu \bar \nu$, 
$\gamma \to \nu \bar \nu$ and $\gamma \gamma \to \nu \bar \nu$ 
are investigated in the presence of a strongly magnetized and dense 
electron-positron plasma.
The amplitudes of the reactions  $\gamma e^{\pm} \to e^{\pm} \nu \bar \nu$ 
and $\gamma \gamma \to \nu \bar \nu$ are obtained.
In the case of a cold degenerate plasma contributions of the considering 
processes to neutrino emissivity are calculated.
It is shown that contribution of the process 
$\gamma \gamma \to \nu \bar \nu$ to neutrino emissivity is 
supressed in comparision with the contributions of the processes 
$\gamma e^{\pm} \to e^{\pm} \nu \bar \nu$ and 
$\gamma \to \nu \bar \nu$.
The constraint on the magnetic field strength in 
the magnetar outer crust is obtained.
\end{abstract}

\section{Introduction}

\indent\indent
Magnetars are highly interesting objects in the our Universe.
Recent 
observations~\cite{Kouveliotou:1998ze,Kouveliotou:1998fd,Gavriil:2002mc,Ibrahim:2002zw} 
give ground to believe that some astrophysical objects (SGR and AXP) are 
magnetars, 
a distinct class of isolated neutron  stars with magnetic field strength of
$B \sim 10^{14}-10^{16}$ G~\cite{Duncan:1992,Duncan:1995,Duncan:1996}, 
i.e. $B \gg B_e$, where 
$B_e = m^2/e \simeq 4.41\times 10^{13}$~G~\footnote{We use natural units
$c = \hbar = k = 1$, $m$ is the electron mass,  $e > 0$ is the elementary
charge.} is the critical magnetic field.
The spectra analysis of these objects is also providing evidence for the
presence of electron-positron plasma in magnetar environment.
In addition, the very density matter is attended in the inner layers both
 the ordinary neutron stars and the magnetars~\cite{Shapiro}. 

The understanding of the important role of quantum processes in the 
magnetar dynamic is the extra stimulus of progress in the astroparticle 
physics. 
It is especially important to investigate
the influence of external field on the quantum processes where
only electrically neutral particles in the initial and the final states
are presented, such as neutrinos and photons. 
The effect of an external field on such processes is associated with two factors. 
First, charged fermions are sensitive to a magnetic field, a major part being 
played here by the electron, since this particle has the maximum specific charge. 
Second, a strong magnetic field has a pronounced effect on the dispersion 
properties of photons and, hence, on the kinematic of processes with its 
participation.

Although the matter in neutron stars is very dense ($\rho \lesssim 10^{11}$ g/cm$^3$ 
in the outer crust), it is fully transparent for neutrino. 
Therefore, it is important to study whole variety of neutrino reactions in 
the young magnetar in order to analyze the cooling.
We will consider the following neutrino reactions in a magnetar crust. 
%They are following:  
\begin{itemize}
\item Pair annihilation process, $e^+ e^- \to \nu \bar \nu$   and  
synchrotron mechanism, 
$e \to e \nu \bar \nu$
are negligible in strongly magnetized, degenerate plasma~\cite{Yakovlev2000}.  
%(D. Yakovlev et al. 2001).  

\item Photoneutrino process, $e \gamma \to  e \nu \bar \nu$. This process
was studied by N.~Itoh et al.~\cite{Itoh1992} in  the plasma without magnetic field
and V.~Skobelev~\cite{Skobelev2000} in the limit of nonrelativistic  degenerate plasma.   
%(N. Itoh et al. 1992 -- in the plasma without magnetic field).

\item Photon conversion process, $\gamma \to \nu \bar \nu$. This process was
studied in the two limits weak magnetic field~\cite{Melrose:1998} and
in strong magnetic field without plasma~\cite{KM_Book}. 
%(D. Melrose et al. 1998 -- in  weak magnetic field, 
%N. Mikheev et al. 1997 -- in strong magnetic field without plasma).

\item Two photon annihilation, $\gamma \gamma \to \nu \bar \nu$. As it was shown in
the recent paper~\cite{RCh05} the amplitude of this process in strongly magnetized, 
degenerate plasma will gain by factor $eB$. 
\end{itemize}

In the present work the process of neutrino cooling
 is investigated in the presence of
strong magnetic field and electron  plasma, when the
magnetic field strength $B$ is the maximal physical parameter,
namely $\sqrt{eB} \gg T,\, \mu \, \omega, \, E$. Here $T$ is the
plasma temperature, $\mu$ is the chemical potential, $\omega$ and
$E$ is the initial photon and electron energies. In
this case almost all electrons and positrons in plasma are on the
ground Landau level.
The more accurate relation for magnetic field
and plasma parameters in this case can be written in the following
form
\begin{eqnarray}
\frac{B^2}{8 \pi} \gg \frac{\pi^2 n_{e^-}^2}{eB} + \frac{eB T^2}{12},
\label{eq:eB}
\end{eqnarray}

\noindent where $n_{e^-}$ are electron  number densities.
In the outer crust of neutron stars the number of  electron density 
%in a strongly magnetized, degenerate plasma 
can be estimated as 
\begin{eqnarray}
n_{e^-} \simeq \frac{m^3}{2\pi^2}\;
\frac{\rho_6\;Z}{A}, \quad \rho_6 = \frac{\rho}{10^6 \mbox{g/cm}^3}\, .
\nonumber
\end{eqnarray}
\noindent For the typically parameters in the outer crust  
($Z=26$, $A=56$ and $T<m$~\cite{Yakovlev2000}), the condition~(\ref{eq:eB}) 
can be performed up to the density 
 $\rho \sim 10^{10}$ g/cm$^3$ for the magnetic field strength 
$B\gtrsim 10^{15}$ G.

Thus, we will investigate the magnetar cooling via neutrino emissivity 
(energy carried out by neutrinos from unit volume per unit time) 
with taking into account of the photon dispersion 
in  strong magnetic field and plasma.

\section{Photon dispersion in the magnetized medium}

\indent\indent
The propagation of the electromagnetic radiation in any active medium is
convenient to describe in terms of normal modes (eigenmodes). In turn, the
polarization and dispersion properties of normal modes are connected with eigenvectors and
eigenvalues of polarization operator correspondingly. In the case of
strongly magnetized plasma 
 in the one loop approximation
the eigenvalues of the
polarization operator can be derived from the previously obtained 
results~\cite{Rojas1979,Rojas1982,Shabad}:

\begin{eqnarray}
\P^{(1)}(q) &\simeq& - \frac{\alpha}{6 \pi}\, 
\left [ q_{\mbox{\tiny $\bot$}}^2 +
\sqrt{q_{\mbox{\tiny $\bot$}}^4 + 
\frac{(6N\omega)^2 q^2}{q_{\mprl}^2}}\, \right ] - q^2\, \Lambda(B) , \label{P1}
\\[2mm]
{\cal P}^{(2)}(q) &\simeq& -\frac{2 eB \alpha}{\pi}\left[
H\left(\frac{q^2_{\mprl}}{4m^2} \right) + {\cal J}(\prl{q}) \right] 
- q^2 \, \Lambda(B), \label{P2} 
\\[2mm]
\P^{(3)}(q) &\simeq&  - \frac{\alpha}{6 \pi}\, 
\left [ q_{\mbox{\tiny $\bot$}}^2  -\sqrt{q_{\mbox{\tiny $\bot$}}^4 + 
\frac{(6N\omega)^2 q^2}{q_{\mprl}^2}}\, \right ] - q^2\, \Lambda(B) , 
\label{P3}
\end{eqnarray}

where
$$
\Lambda(B) = \frac{\alpha}{3 \pi}\,\left[1.792 - \ln
(B/B_e)\right], \quad 
N = \int \limits_{-\infty}^{+\infty} dp_z \,
\left [f_{-}(E) - f_{+}(E)\right ],
$$
\beq
\nonumber
{\cal J}(\prl{q}) = 2 \prl{q}^2 m^2 \int \frac{dp_z}{E} \,
\frac{f_{-}(E) + f_{+}(E)}
{(\prl{q}^2)^2 - 4 \prl{(pq)}^2}\,, \qquad E = \sqrt{p_z^2 + m^2},
\eeq
$f_{\pm}(E) \, = \, [e^{(E \, \pm \, \mu)/T} \, + \, 1]^{-1}$
 are the electron (positron) distribution functions,
\beq
\label{eq:H0}
%\nonumber
&&H(z)=\frac{1}{\sqrt{z(1 - z)}} \arctan \sqrt{\frac{z}{1 - z}} - 1,
\quad 0 \leqslant z \leqslant 1,
\\ [3mm]
\label{eq:H1}
&&H(z) = - \frac{1}{2\sqrt{z(z-1)}}
\ln  \frac{\sqrt{z} + \sqrt{z-1}}{\sqrt{z} - \sqrt{z-1}}  - 1 + 
\,\frac{i\pi}{2\sqrt{z(z-1)}}, \quad z > 1.
\eeq

\noindent 
In the case of strongly degenerate plasma $(T \ll \mu, \, m)$ one can obtained 
the analytical expressions for ${\cal J}(q_{\mprl})$ and $N$ integrals: 
%The integrals ${\cal J}(q_{\mprl})$ and $N$ 
%can be calculated in the case  of a strongly 
%degenerate plasma $(T \ll \mu, \, m)$ and we obtained the following result
%
\begin{eqnarray}
\label{eq:J0}
%\nonumber
&&{\cal J}(q_{\mprl}) = - \frac{1}{2\sqrt{ z(1 - z)}}
\bigg (  \arctan  \bigg [ \frac{v_F - v_\phi + z v_F (v_\phi^2 -1)}
{(v_\phi^2 -1) \sqrt{z(1 - z)}}  \bigg] +
%\nonumber
\\[3mm]
&& + \arctan  \bigg [ \frac{v_F + v_\phi + z v_F (v_\phi^2 -1)}
{(v_\phi^2 -1) \sqrt{z(1-z)}}  \bigg] \bigg ), \qquad 0 \leqslant z \leqslant 1,
\nonumber
\eeq
%\\[3mm]
%
\beq
\label{eq:J1}
&&{\cal J}(q_{\mprl}) = - \frac{1}{4\sqrt{ z(z - 1)}}
\bigg (  \ln  \bigg [ \frac{v_F - v_\phi + (v_\phi^2 -1)(z v_F - \sqrt{ z(z - 1)}\,)}
{v_F - v_\phi + (v_\phi^2 -1)(z v_F + \sqrt{ z(z - 1)}\,)}  \bigg] +
%\nonumber
\\[3mm]
&& + \ln  \bigg [ \frac{v_F + v_\phi + (v_\phi^2 -1)(z v_F - \sqrt{ z(z - 1)}\,)}
{v_F + v_\phi + (v_\phi^2 -1)(z v_F + \sqrt{ z(z - 1)}\,)}  \bigg]\bigg ) 
-\frac{i\pi \theta(v_F|v_\phi| - 1)}{2\sqrt{z(z-1)}}\, , \qquad z>1,
\nonumber
\\[3mm]
&&z = \frac{q_{\mprl}^2}{4m^2}\, , \quad 
v_F = \frac{\sqrt{\mu^2 - m^2}}{\mu}, \quad v_\phi = 
\frac{\omega}{q_z}, \quad N = 2p_F = 2\sqrt{\mu^2 - m^2}.
\nonumber
\end{eqnarray}

\noindent Here the four-vectors with indices $\bot$ and $\parallel$
belong to the Euclidean \{1, 2\}-subspace and the Minkowski
\{0, 3\}-subspace correspondingly in the frame were the magnetic
field is directed along $z$ (third) axis; 
$(ab)_{\mprp} = (a \Lambda b) = a_\alpha \Lambda_{\alpha \beta} b_\beta$, 
$(ab)_{\mprl} = (a \tilde \Lambda b) = a_\alpha \tilde \Lambda_{\alpha \beta} b_\beta$,   
where the tensors
$\Lambda_{\alpha \beta} = (\varphi \varphi)_{\alpha \beta}$,\,
$\widetilde \Lambda_{\alpha \beta} =
(\tilde \varphi \tilde \varphi)_{\alpha \beta}$, with equation
$\widetilde \Lambda_{\alpha \beta} - \Lambda_{\alpha \beta} =
g_{\alpha \beta} = diag(1, -1, -1, -1)$ are introduced.
$\varphi_{\alpha \beta} =  F_{\alpha
\beta} /B$ and
${\tilde \varphi}_{\alpha \beta} = \frac{1}{2} \varepsilon_{\alpha \beta
\mu \nu} \varphi_{\mu \nu}$ are the dimensionless field tensor and dual
field tensor correspondingly.
\begin{figure}[h]
\centerline{\includegraphics{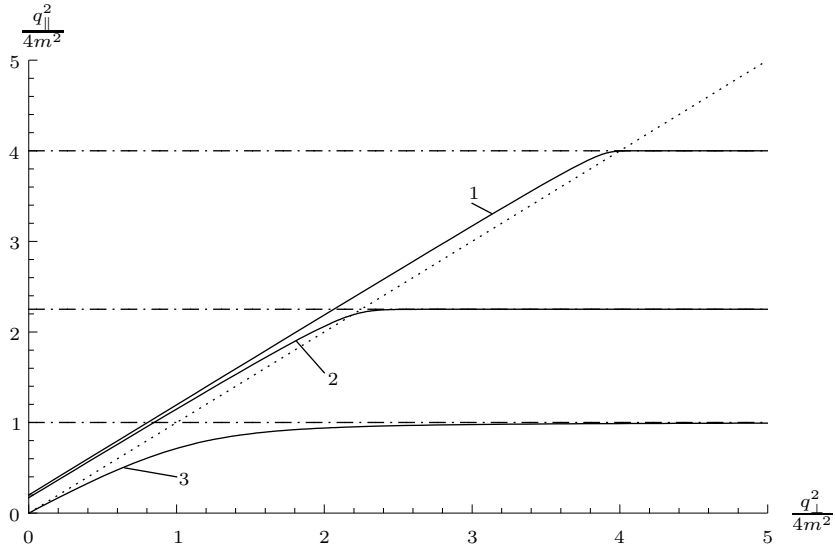}}
\caption{Photon dispersion in a strong magnetic field ($B/B_e = 200$) 
and degenerate plasma vs. chemical potential $\mu = 1$ MeV -- 1, 
$\mu = 0.75$ MeV -- 2 and  without plasma -- 3.
Dotted line corresponds to the vacuum dispersion law, $q^2 = 0$. The angle 
between the photon momentum and the magnetic
field direction is $\pi/2$. }
\label{fig:dis1}
\end{figure}
%%%%%%%%%%%%%%%%%%%%%%%%%%%%%%%%%%%%%%%%%%%%%%%%%%%%%%%%%%%%%%%%%%%%%%%

The dispersion properties of the normal modes could be defined from the
dispersion equations
\beq
q^2 - \P^{(\lm)}(q) = 0 \qquad (\lm = 1, 2, 3).
\label{disper}
\eeq
Their analysis shows that  1 and 2  modes with polarization vectors
\beq
\ee_\alpha^{(1)}(q) = \frac{(q \varphi)_\alpha}{\sqrt{q_{\mprp}^2}},
\qquad
\ee_\alpha^{(2)}(q) = \frac{(q \tilde \varphi)_\alpha}
{\sqrt{q_{\mprl}^2}}.
\label{epsilon}
\eeq
are only physical ones in the case under consideration, just as it is in the
pure magnetic field~\footnote{ Symbols 1 and 2 correspond to
the $\|$ and $\perp$ polarizations in pure magnetic
field~\cite{Adler:1971} and $E$- and $O$- modes in magnetized
plasma~\cite{Duncan:1995}.}. However, it should be emphasized  that this 
coincidence is
approximate to within $O(1/\beta)$ and $O(\alpha^2)$ accuracy.

Notice, that in plasma only the eigenvalue $\P^{(2)}(q)$ is modified in
comparison with pure magnetic field case. It means that  the dispersion
law of the mode 1 is the same one as in the magnetized vacuum, where its
deviation from the vacuum law, $q^2 = 0$, is negligibly small.
From the other hand, the dispersion properties  of the mode 2 essentially differ from the
magnetized vacuum ones. In the Fig.~\ref{fig:dis1}
  the photon dispersion in both
strong magnetic field and magnetized degenerate plasma are depicted at chemical potential.  
One can see that in the presence of the magnetized plasma there exist the kinematical 
region, where $q^2 >0$ (at $q_{\mprl}^2 < 4m^2$) contrary 
to the case of  pure magnetic field. This fact could lead to the 
modification of the kinematics of 
the different photon-neutrino processes.   
For example, the photon conversion $\gamma \to \nu \bar \nu$  
forbidden in the magnetic field without plasma becomes allowed in 
this region~\cite{RCh05}.
It is connected with the 
appearance of the plasma frequency
$\omega_{pl}^2 = 2\alpha eB v_F/\pi.$

Moreover, as can be seen from the Fig.~\ref{fig:dis1} in degenerate plasma there 
is to be the shift of the $e^+e^-$ 
pair-creation threshold which in pure magnetic field is defined by the relation 
$q^2_{\mprl} = 4m^2$. One can see that in the region $|v_\phi| > 1/v_F$ ($|q_z| < 2p_F$)
the last terms in ~(\ref{eq:H1}) and (\ref{eq:J1}) cancel each other and the only 
contribution    to the imaginary part of $\P^{(2)}(q)$ comes from the logarithm function
in (\ref{eq:J1}). It is the fact that leads to the shift of the pair creation threshold
from $q^2_{\mprl} = 4m^2$ to
\beq
q^2_{\mprl} =
2\left (\mu^2 - p_F|q_z| + \mu \, \sqrt{(p_F - |q_z|)^2 + m^2}\right ).
\label{eq:shift}
\eeq

\noindent This result is in agreement with simple kinematical analysis of the process
 $\gamma_2 \to e^+ e^-$   in degenerate plasma. Indeed, using the energy and 
 momentum conservation laws with obviuos conditions 
 $E \geqslant \mu$ and $|p_z| \geqslant p_F$ for the electron we come 
 to the result~(\ref{eq:shift}).

\section{Neutrino emissivity}

\indent\indent
Our main goal is  to obtain the neutrino emissivity in various neutrino reactions. 
A general expression for neutrino emissivity 
can be defined in the following way:

\begin{eqnarray}
Q = \frac{1}{V}\;\int \prod \limits_{i} \D \Gamma_i f_i \,
\prod \limits_{f} \D \Gamma_f (1\pm f_f) \; q_0\,
\frac{|S_{if}|^2}{\tau},
\label{eq:Q}
\end{eqnarray}
\noindent 
where  $\D \Gamma_i$ $(\D \Gamma_f)$ are the number of states of 
initial (final) particles; 
$f_i$ $(f_f)$ are the corresponding of distribution functions, 
the sign $+$ $(-)$ corresponds  to final bosons (fermions); 
$q_0$ is the neutrino pair energy;
$V$ is the plasma volume, $\tau$ is the interaction time, 
$S_{if}$ is the $S$ - matrix element. 

We will consider the case of relatively low momentum transfers 
 $|q^2| \ll m_W^2$ under calculation of the $S$ - matrix elements.  
Under this condition, the weak interaction of neutrinos with electrons can 
be considered  
in the local limit by using the effective Lagrangian

\begin{eqnarray}
{\cal L} \, = \, \frac{G_F}{\sqrt 2}\,
\big [ \bar e \gamma_{\alpha} (C_V + C_A \gamma_5) e \big ] \,
j_{\alpha} \,,
\label{eq:L}
\end{eqnarray}
\noindent where $C_V = \pm 1/2 + 2 \sin^2 \theta_W, \, C_A = \pm 1/2$,
$j_{\alpha} = \bar \nu \gamma_{\alpha} (1+ \gamma_5) \nu$ -- is the neutrino
current. 

The analysis shows, that the integral over 
phase space in~(\ref{eq:Q}) gains its value in the vicinity of  
the plasma frequency. In this region of 
the dispersion law for a mode-2 photon can be written as
$\omega^2 = q_{\mprp}^2 + q_z^2 + \omega_{pl}^2$. 
Using the approximation of dispersion law we have obtain the simple expressions 
for neutrino emissivity due the  processes 
$e\gamma \to e\nu \bar \nu$, 
$\gamma \to \nu \bar \nu$ and $\gamma \gamma \to \nu \bar \nu$ 
 in the  two cases of 
nonrelativistic and relativistic plasma.

The emissivity due to the photoneutrino  process is
\begin{eqnarray}
%\nonumber
Q_{\gamma e \to e \nu \bar \nu} \simeq 
 \; 1.3\times 10^{19} \frac{\mbox{erg}}{\mbox{cm}^3 \; \mbox{s}} \;
\frac{B}{B_e}\; \left (\frac{T}{m}\right )^8 \; 
\frac{T}{p_F} , \quad \mu \sim m ;  
\end{eqnarray}
\begin{eqnarray}
\nonumber 
Q_{\gamma e \to e \nu \bar \nu} \simeq 5.4 \times 10^{17} 
\; \frac{\mbox{erg}}{\mbox{cm}^3 \; \mbox{s}}\;
\frac{B}{B_e}\; 
\left (\frac{\mu}{m}\right )^5 \;\left(\frac{T}{\omega_{pl}} \right)^{3/2} 
\left(\frac{\omega_{pl}}{2\mu} + 1 \right) \times
\\
%\nonumber
\times \int \limits_0^1 dx\; (1-x) \frac{(\omega_{pl}/2\mu)^2-x^2}
{1-\exp{\left [-\frac{\mu}{T}(\omega_{pl}/2\mu-x)\right ]}}, \quad \mu \gg m. 
\end{eqnarray}

The emissivity due to the photon conversion  process is

\begin{eqnarray}
%\nonumber
Q_{\gamma \to \nu \bar \nu} \simeq 10^{21} \; \frac{\mbox{erg}}{\mbox{cm}^3 \; \mbox{s}}\;
\left (\frac{T}{m}\right )^9 \,
\left(\frac{\omega_{pl}}{T}\right)^4
\, \left [18.7 + 3.3 \,\left(\frac{\omega_{pl}}{T}\right )^2 \right ] , \quad \mu \sim m;
\end{eqnarray}
\begin{eqnarray}
%\nonumber
Q_{\gamma \to \nu \bar \nu} \simeq 
10^{20} \; \frac{\mbox{erg}}{\mbox{cm}^3 \; \mbox{s}}\;
\left (\frac{T}{m}\right )^9 \,
\left(\frac{\omega_{pl}}{T}\right)^{15/2}
\, \left [5.5 + 9.0 \,\frac{T}{\omega_{pl}} \right ] 
\exp{\left(-\frac{\omega_{pl}}{T}\right)}, 
\\
\nonumber
\quad \mu \gg m. 
\end{eqnarray}

The emissivity due to the two photon annihilation  process is

\begin{eqnarray}
%\nonumber
Q_{\gamma \gamma \to \nu \bar \nu} \simeq 
5.3 \times 10^{19} \; \frac{\mbox{erg}}{\mbox{cm}^3 \; \mbox{s}}\; 
 \left (\frac{\omega_{pl}}{T}\right )^4 \,
\left (\frac{T}{m}\right )^{11} \, , \quad  \mu \sim m\, ,
\end{eqnarray}

\begin{eqnarray}
\nonumber
Q_{\gamma \gamma \to \nu \bar \nu} &\simeq& 
10^{16} \; \frac{\mbox{erg}}{\mbox{cm}^3 \; \mbox{s}}\; 
\left (\frac{B}{B_e}\right )^2 \,
\left (\frac{T}{m}\right )^7 \,
\left(\frac{\omega_{pl}}{T}\right)^3\; \left(\frac{m}{\mu}\right)^6 \times
\\ [2mm]
%\nonumber  
&\times& \left [2.5 + 
2.0\left(\frac{T}{\omega_{pl}}\right) \right ] 
\exp{\left(-\frac{2\omega_{pl}}{T}\right)}\; , \quad 
\mu \gg m \; . 
\end{eqnarray}
\begin{figure}[h]
\centerline{\psfig{file=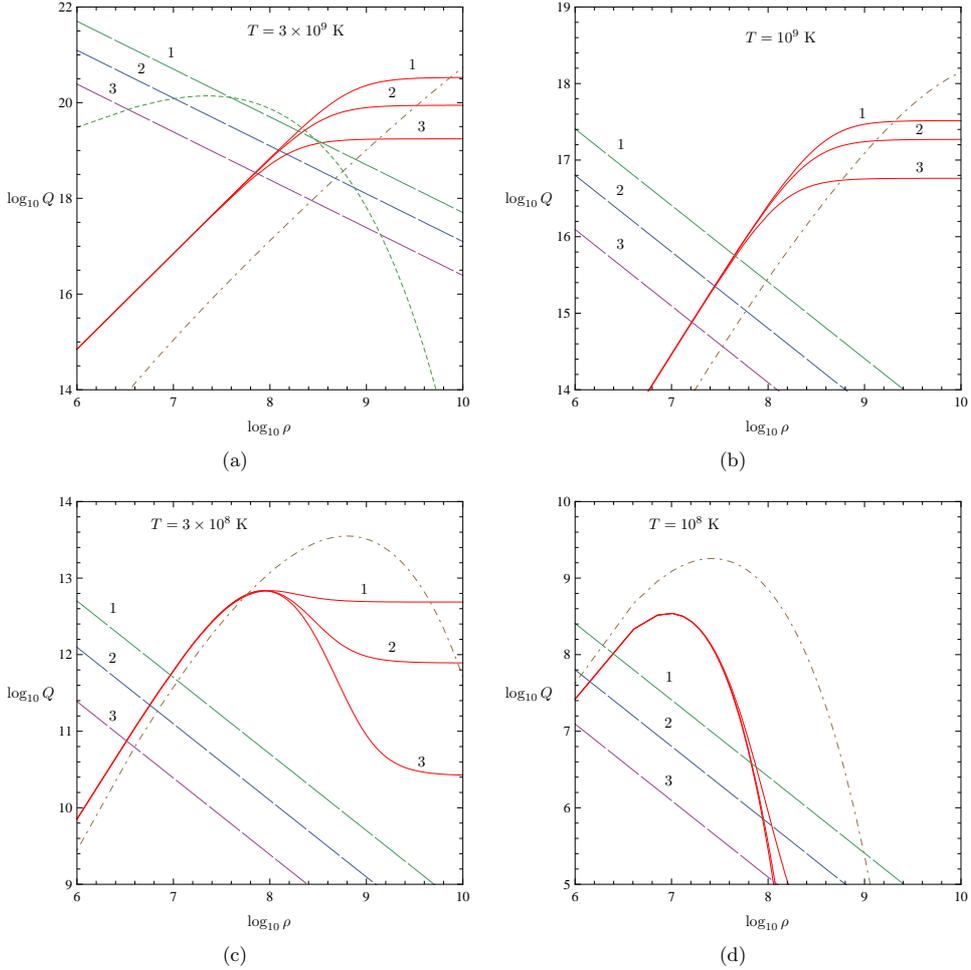,width=5in}}
%\centerline{\includegraphics{fig2.eps}}
\caption{The dependence of the contributions in the neutrino 
emissivity (erg$\cdot$cm$^{-3}$ $\cdot$s$^{-1}$) on matter density
 (g$\cdot$cm$^{-3}$) in the outer crust  of magnetized neutron star
for various temperatures  $T = 3\times 10^9$~K (a),  
$T = 10^9$ K (b), $T = 3\times 10^8$~K (c),    
$T = 10^8$ K (d) and magnetic field values 
1 -- $10^{16}$ G., 
2 -- $5 \times 10^{15}$ G., 3 -- $2.2 \times 10^{15}$ G. 
The solid line corresponds to the $\gamma \to \nu \bar \nu$ process. 
The dashed line corresponds to the photoneutrino process.
 The dotted line corresponds to the $e^+e^-$ pair annihilation process 
at $B=0$~\cite{Yakovlev2000}.
The chain line corresponds to the 
plasmon decay process~\cite{Yakovlev2000}. 
Pair annihilation process become negligible at $T \lesssim 10^9$ K.}
\label{fig:fig2}
\end{figure}
%%%%%%%%%%%%%%%%%%%%%%%%%%%%%%%%%%%%%%%%%%%%%%%%%%%%%%%%%%%%%%%%%%%%%%%

We analyse the $\gamma \gamma \to \nu \bar \nu$ process contribution in the 
neutrino emissivity in the regions 
of the temperature ($10^8 \lesssim T \lesssim 3\times 10^{9}$~K), the density 
($10^6 \lesssim \rho \lesssim 10^{10}$~g/cm$^3$) and the magnetic field 
strength ($B \lesssim 10^{16}$~G.).
The obtaining results show, that the 
 influence of this process on the 
 emissivity is suppressed as compared with the contributions of photoneutrino and 
photon conversion processes under these conditions. 
 Therefore, the possible influence of process $\gamma \gamma \to \nu \bar \nu$ 
on the magnetar cooling is negligible.

\section{Application to magnetar cooling}

\indent\indent
We have made the numerical calculation of the neutrino emissivity dependence 
on the density caused by 
processes $\gamma e \to e \nu \bar \nu$ (dashed line) and 
$\gamma \to \nu \bar \nu$ (solid line).
The results are represented in Figures~\ref{fig:fig2} (a -- d) 
for different temperatures 
$T=3\times 10^9,\,10^9,\, 3\times 10^8, \,10^8$~K and magnetic field strength
a -- $10^{16}$ G., b -- $5 \times 10^{15}$ G., c -- $2.2 \times 10^{15}$~G.
The dotted line corresponds to the $e^+e^-$ pair 
annihilation process (Fig.~\ref{fig:fig2}a).
The chain line corresponds to the plasmon decay 
process (Fig.~\ref{fig:fig2} (a -- d)).
%~\footnote{We use the data from~\cite{Yakovlev2000}}.

As can be seen from Fig.~\ref{fig:fig2} (a -- c) the photoneutrino process is provided 
the leading contribution in the neutrino emissivity 
in the density region $10^6 \lesssim \rho \lesssim 10^8$ g/cm$^3$ at 
the temperature $T=3\times 10^9,\,10^9$~K and in the density 
region $10^6 \lesssim \rho \lesssim 10^7$ g/cm$^3$ at 
the temperature $T=3\times 10^8$~K.
 The photon conversion process are suppressed by plasma frequency
in this region. % $\rho \lesssim 10^{10}$ g/cm$^3$ 
 On the other hand, it is provided 
the main contribution in the neutrino emissivity 
in the density region $10^8 \lesssim \rho \lesssim 10^{10}$ g/cm$^3$.
It is worth noting that the our results are unsuitable 
for the density $\rho \gtrsim 10^{10}$ g/cm$^3$.

As the application, we will consider the neutron stars cooling model~\cite{Yakovlev2000}. 
 The neutrino processes in the neutron stars crust are provided of the neutrino cooling 
during $10^{-2}\lesssim t \lesssim 100$ years in this model. 
In addition, the plasmon decay in a weakly magnetized plasma is dominant  in this 
period.
On the other hand (see Fig. 2a and 2b), at the magnetic field strength 
$B \gtrsim 5\times 10^{15}$~G 
and the temperature $10^9 \lesssim T \lesssim 3\times 10^9$~K 
 both processes 
$\gamma e \to e \nu \bar \nu$ and $\gamma \to \nu \bar \nu$ are  leading 
 as compared with the plasmon decay in a weakly magnetized plasma. 
%We will of opinion that
Let us discuss the possible consequences of our results.

\begin{itemize}

\item Assumig, that the temperature profile in the outer crust weakly depends 
on magnetic field strength.

\item Let us assume also that the magnetar cooling regime at the time 
$t \gtrsim 10^{3}$ years is the some one as for the ordinary neutron stars. 

\end{itemize}

When we can obtain the upper limit for magnetic field value 
$5\times 10^{15}$~G. However, it is rough estimation of magnetic field. 

In conclusion, we have consider the influence of a strongly magnetized 
plasma  on the photon-neutrino 
processes $\gamma e^{\pm} \to e^{\pm} \nu \bar \nu$, 
$\gamma  \to  \nu \bar \nu$ and $\gamma \gamma \to  \nu \bar \nu$.
 The changes of the photon dispersion properties in a magnetized 
medium are investigated. 
 We have obtained the simple expressions for neutrino emissivity in 
the cold plasma limit. These results can be used for the simulation of
the magnetar cooling. 
It is shown, that the possible influence of process 
$\gamma \gamma \to \nu \bar \nu$ 
on the magnetar cooling is negligible in the regions 
of temperature ($10^8 \lesssim T \lesssim 3\times 10^{9}$ K), density 
($10^6 \lesssim \rho \lesssim 10^{10}$ g/cm$^3$) and magnetic field 
strength ($B \lesssim 10^{16}$ G.).
From the possible modification of the magnetar cooling  scenario 
we have obtained the upper bound on the magnetic field strength
$B \lesssim 5 \times 10^{15}$ G. 

\bigskip

{\bf Acknowledgements}  

We express our deep gratitude to the organizers of the 
Seminar ``Quarks-2008'' for warm hospitality.

The work was supported in part by the Russian Foundation for Basic Research
under the Grant No.~07-02-00285-a, and by the Council on Grants by the
President
of the Russian Federation for the Support of Young Russian Scientists
and Leading Scientific Schools of Russian Federation under the Grant
No.~NSh-497.2008.2.

%%%%%%%%%%%%%%%%%%%%%%%%%%%%%%%%%%%%%%%%%%%%%%%%%%%%%%%%%%%%%%%%%%%%%%%%%
%\newpage

%%%%%%%%%%%%%%%%%%%%%%%%%%%%%%%%%%%%%%%%%%%%%%%%%%%%%%%%%%%%%%%%%%%%%%%%%

\end{document}